\documentstyle[epsf,seceq,preprint]{ptptex}

\def\be{\begin{equation}}
\def\ee{\end{equation}}
\def\bea{\begin{eqnarray}}
\def\eea{\end{eqnarray}}

\def\simm#1{\mathop{\vtop{\ialign{##\crcr
        $\hfil\displaystyle{#1}\hfil$\crcr\noalign{\kern0.5pt\nointerlineskip}
        $\sim$\crcr\noalign{\kern0.5pt}}}}\limits}

\preprintnumber[4cm]{%<-- [..]: optional width of preprint # column.
UTCCP-P-35\\ April 1998}

\title{%        %You can use \\ for explicit line-break
Quantum Chromodynamics with Many Flavors%
\thanks{Talk presented by K.~Kanaya
at the 1997 Yukawa International Seminar (YKIS'97) on 
``Non-Perturbative QCD --- Structure of the QCD Vacuum ---'',
YITP, Kyoto, Japan, 2--12 Dec.\ 1997. To be published in the proceedings
[Prog.\ Theor.\ Phys.\ Suppl.].
}
}

\author{%       %Use \sc for the family name
Y.~{\sc Iwasaki}, K.~{\sc Kanaya}, 
S.~{\sc Kaya},$^*$ S.~{\sc Sakai},$^{\dag}$ 
and T.~{\sc Yoshi\'e}
}

\inst{%         %Affiliation, neglected when [addenda] or [errata]
Center for Computational Physics and Institute of Physics,\\ 
University of Tsukuba, Tsukuba, Ibaraki 305-8577, Japan\\
\vspace{2mm}
$^*$Institute of Particle and Nuclear Studies,\\
High Energy Accelerator Research Organization (KEK),\\
Tsukuba, Ibaraki 305-0801, Japan\\ 
\vspace{2mm}
$^{\dag}$Faculty of Education, Yamagata University,\\
Yamagata 990-8560, Japan
}

\recdate{%      %Editorial Office will fill in this.
\today
}

\abst{
We investigate the phase structure of lattice QCD for
general number of flavors $N_F$.
Based on numerical data combined with the results of the
perturbation theory we propose the following picture:
When $N_F \ge 17$, there is only 
one IR fixed point at vanishing gauge coupling, i.e., 
the theory in the continuum limit is trivial. 
On the other hand, when $16 \ge N_F \ge 7$,
there is a non-trivial fixed point.
Therefore, the theory is non-trivial with anomalous dimensions, 
however, without quark confinement.
Theories which satisfy both quark confinement and spontaneous
chiral symmetry breaking in the continuum limit
exist only for $N_F \le 6$.
}

\begin{document}

\maketitle
  
\section{Introduction}
\label{sect:intro}

At low temperatures, QCD with a small number of light quarks 
is known to have three 
characteristic properties of asymptotic freedom, quark confinement 
and spontaneous breakdown of chiral symmetry.
Among them, asymptotic freedom is reliably investigated 
by the perturbation theory.
It is well-known that when the number of flavors $N_F$ exceeds 
$16 \frac{1}{2}$, the asymptotic freedom is lost.
Then, a question which naturally arises is whether
there is a constraint on $N_F$ for
quark confinement and/or the spontaneous breakdown of chiral symmetry.
Because these properties are essentially non-perturbative in nature, 
we have to carry out a non-perturbative investigation.

In this paper, we study the $N_F$ dependence of the QCD vacuum
by numerical simulations on the lattice. 
Because the Wilson fermion formalism is the only formalism for lattice
quarks that preserves manifest flavor symmetry, 
we use the Wilson fermion formalism in our study of $N_F$ dependence
in QCD.
Based on the numerical data obtained for $N_F$ from 2 up to 300, 
we elucidate the phase structure of QCD as a function of $N_F$.
We conjecture the following picture:
When $N_F \ge 17$, there is only
one IR fixed point at vanishing gauge coupling. 
Therefore, the theory in the continuum limit
is trivial. On the other hand, when $16 \ge N_F \ge 7$, 
there is a non-trivial IR fixed point.
Therefore, for this case, the theory is non-trivial with anomalous 
dimensions, however, without quark confinement.
Theories which satisfy both quark confinement and spontaneous
chiral symmetry breaking in the continuum limit
exist only for $N_F \le 6$.
Our results are in part reported in Refs.~\citen{lat94} and \citen{lat96}.

This paper is organized as follows.
Before going into details of our results obtained from numerical 
simulations, in Sec.~\ref{sec:betafunc}, 
we first summarize results and speculations from a perturbative study 
of the renormalization group beta function,
and then present our conjecture for the $N_F$ dependence of the 
beta function, based on our numerical data obtained on the lattice.
In Sec.~\ref{sec:model}, 
our lattice model and simulation parameters are described.
In order to investigate the $N_F$ dependence of the QCD vacuum 
in the continuum limit,
we study the phase structure of lattice QCD at zero temperature 
for general number of flavors, in particular, for $N_F \ge 7$. 
When the phase diagram becomes clear, 
we are able to see the nature of the QCD vacuum in the continuum limit,
and eventually answer the question about the condition on $N_F$ 
for confinement and spontaneous breakdown of the chiral symmetry.
Results for the phase structure in the strong coupling limit are given 
in Sec.~\ref{sec:strong}.
Sec.~\ref{sec:finitebeta} is devoted to the phase structure
off the strong coupling limit.
We then discuss the nature of the deconfining phase in 
Sec.~\ref{sec:deconf}.
Our conclusions and conjectures concerning the phase structure
of QCD for general number of flavors are summarized in 
Sec.~\ref{sec:conclusions}.

\section{Beta function}
\label{sec:betafunc}

The perturbative beta function of QCD is universal up to two-loop
order:
$%\be
\tilde{\beta}(g) = - b_0 g^3 - b_1 g^5 + \cdots.
$%\ee
where
$
b_0 = \frac{1}{16\pi^2} \left( 11-\frac{2}{3}N_F \right)
$
and
$
b_1 = \frac{1}{(16\pi^2)^2} \left( 102-\frac{38}{3}N_F \right).
$
The coefficient $b_0$ changes its sign at $N_F = 16\frac{1}{2}$.
Therefore, for $N_F \geq 17$, 
asymptotic freedom is lost and the IR limit is free, 
i.e.\ we have no confinement.
It is well-known that the second coefficient $b_1$ changes its sign 
already at $N_F \approx 8.05$.
The values of the two coefficients imply that
for $N_F \sim 9$, the IR fixed point appears at $g \sim 3$.
However, this value of the gauge coupling constant is too large
to trust the result from the perturbation theory.
In contrast to this case, for $N_F \sim 16$, the IR fixed point appears in a 
perturbative region. 
Therefore, it is plausible that the beta function has a non-trivial 
IR fixed point for $N' \leq N_F \leq 16$ with some $N' \le 16$.
When such an IR fixed point exists, the coupling constant cannot become 
arbitrarily large in the IR region --- 
this implies that quarks are not confined.

\begin{figure}[tb]
\centerline{
a)\epsfxsize=5.2cm\epsfbox{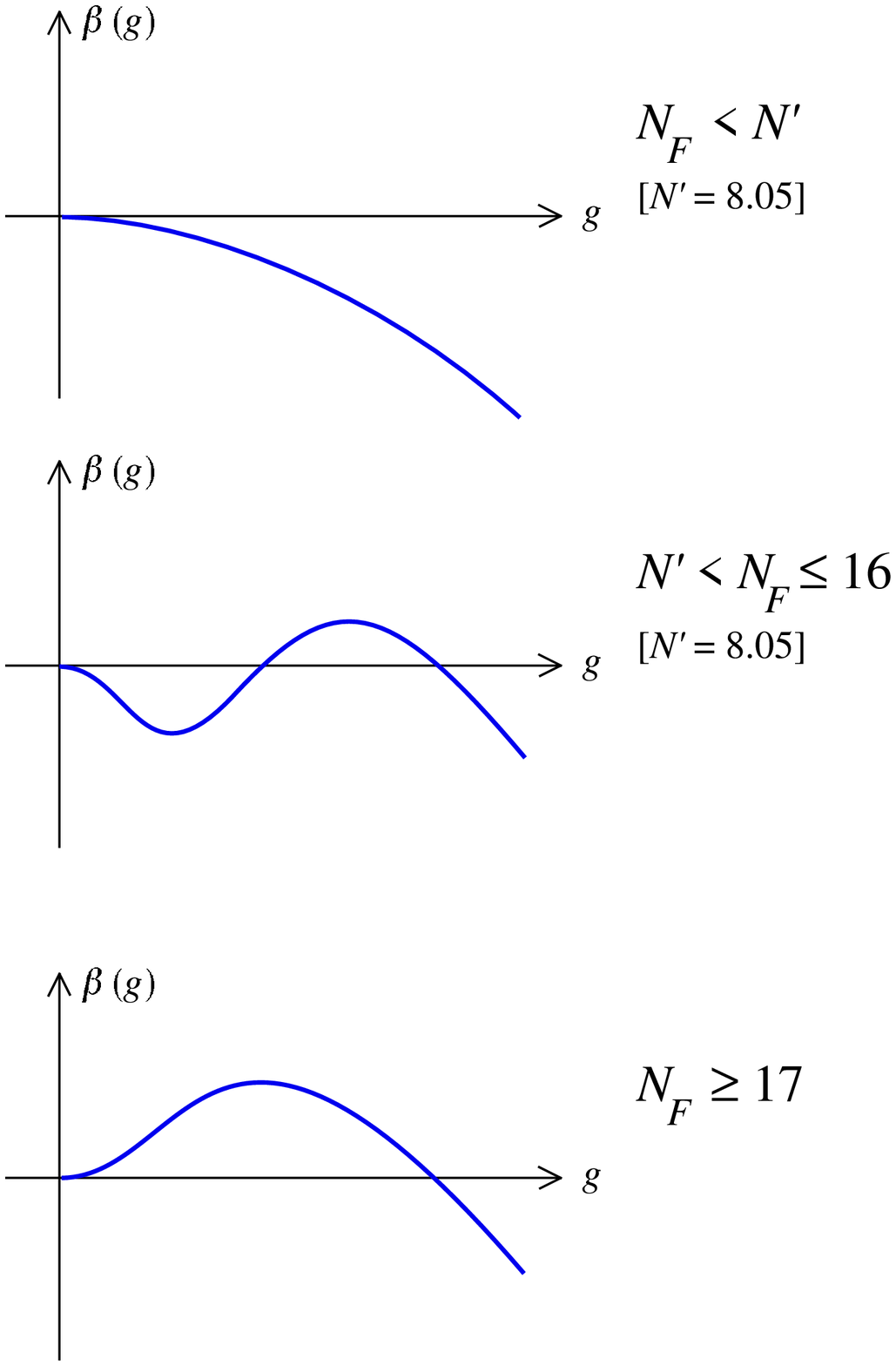}
\makebox[1cm]{}
b)\epsfxsize=5.2cm\epsfbox{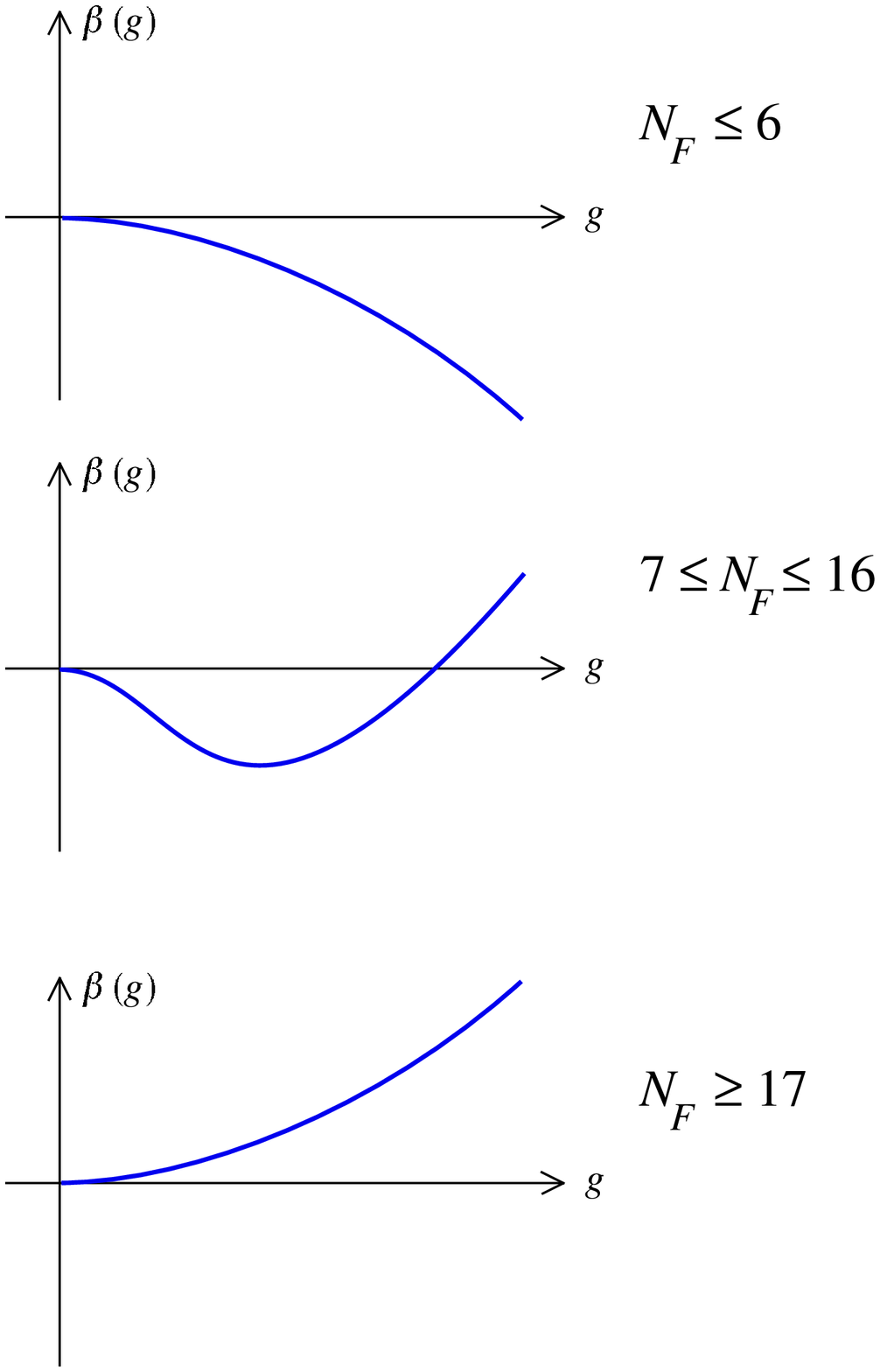}
}
\caption{Renormalization group beta function.
(a) Conjecture by Banks and Zaks \protect\cite{Banks1} 
assuming confinement in the strong coupling limit for all $N_F$.
(b) Our conjecture deduced from the results of lattice simulations.
}
\label{BetaFunc}
\end{figure}

A pioneering study on the $N_F$ dependence of the QCD vacuum was
made by Banks and Zaks in 1982 \cite{Banks1}.
Based on an early result obtained on the lattice \cite{Kogut79},
they assumed that the beta function is negative in the strong 
coupling limit for all $N_F$. 
Using the perturbative results for $N' \leq N_F \leq 16$ mentioned above, 
they conjectured Fig.~\ref{BetaFunc}(a) as the simplest $N_F$ dependence
of the beta function, 
and studied the phase structure of QCD based on this $N_F$
dependence of the beta function.%
\footnote{
For different approaches, see Refs.~\citen{Oehme,Nishijima}
and \citen{Appelquist}.
When the quarks are confined, the sigma model might be a
reasonable effective theory of QCD.  The $N_F$-dependence
of the vacuum in a sigma model is studied in Ref.~\citen{sigma}.
}
Because of an additional non-trivial UV fixed point for $N_F \ge N'$,
their conjecture for the phase structure is complicated.

The assumption of confinement in the strong coupling limit plays 
an essential role in the arguments of Banks and Zaks.
Here, a quite important fact to be noted is that the lattice study
cited by them
is a study of pure gauge theory without quarks \cite{Kogut79}.

When dynamical quarks are coupled, the vacuum structure become 
much complicated. In this case,
we can formally rewrite the theory in terms of mesonic and baryonic
fields [see, for example, Ref.~\citen{SCENc}].
When mesons and baryons are physically relevant dynamical variables,
i.e.\ when quarks are confined,
the effective action for hadrons is useful 
to study the phase structure of the system. 
Oppositely, when the resulting effective action for hadrons 
is a well-defined action of weakly interacting particles,
it is plausible that quarks are confined.
In the strong coupling limit,
computation of the effective theory for hadrons
are carried out either in the large $N_c$ limit \cite{SCENc},
using a meanfield approximation ($1/d$ expansion) \cite{SCEmf,SCEmfMq},
or using a heavy quark mass expansion \cite{SCEmfMq,SCEMq}.
In these cases, the effective action for mesons and baryons leads 
to a world with spontaneously broken chiral symmetry.
Here, in contrast to these cases,
we are interested in the case where $N_c=3$ and quarks are light.
In this case, the effective action is quite complicated
for general $N_F$;
diagrams with quark loops, which are suppressed at large $N_c$,
large $d$ or large $m_q$, become important.
Therefore, in the realistic cases, 
it is difficult to analytically deduce the vacuum structure 
as a function of $N_F$ even in the strong coupling limit.

Therefore, in our previous work \cite{previo}, 
we performed a series of numerical simulations of QCD 
in the strong coupling limit for various $N_F$,
using the Wilson fermion formalism for lattice quarks. 
We found that, when $N_F \ge 7$, 
light quarks are deconfined and chiral
symmetry is restored at zero temperature 
even in the strong coupling limit.

We extend this study to weaker couplings and to larger $N_F$. 
Based on the numerical results obtained on the lattice 
combined with the results of the perturbation theory 
(see the following sections for details), 
we conjecture Fig.~\ref{BetaFunc}(b) for the $N_F$ dependence of the
beta function:
When $N_F$ is smaller than a critical value, 
the beta function is negative for all values of $g$.
Quarks are confined and the chiral symmetry is spontaneously broken
at zero temperature.
We find that the critical number of flavors is 6.
(Corresponding critical $N_F$ is 2 for the case $N_c=2$ \cite{lat94}.)
On the other hand, when $N_F$ is equal or larger than 17,
we conjecture that the beta function is positive for all $g$,
in contrast to 
the conjecture by Banks and Zaks shown in Fig.~\ref{BetaFunc}(a).
The theory is trivial in this case.
When $N_F$ is between 7 and 16, the beta function changes 
sign from negative to positive with increasing $g$. 
Therefore, the theory has a non-trivial IR fixed point. 

\section{Model and simulation parameters}
\label{sec:model}

Lattice QCD is defined on a hypercubic lattice in 4-dimensional 
Euclidian space with lattice spacing $a$. 
Formulation of lattice QCD is described in Ref.~\citen{KanayaYKIS97}.

In this study, we use the standard one plaquette gauge action
combined with the standard Wilson quark action:
$ S = S_{gauge} + S_{quark}$ with
\bea
S_{gauge} & = & - \frac{\beta}{6} \sum_{x,\mu\neq\nu}{\rm Tr}
\left( U_{x,\mu}U_{x+\hat\mu,\nu}
U_{x+\hat\mu+\hat\nu,\mu}^{\dag} U_{x+\hat\nu,\nu}^{\dag}\right),
\label{eqn:gaction} \\
S_{quark} & = & \sum_{f,x} \left[ \bar{\psi}_x^f \psi_x^f
-K\left\{\bar{\psi}_{x}^f(1\!-\!\gamma_{\mu})U_{x,\mu}{\psi_{x+\hat{\mu}}^f}
+\bar{\psi}_{x+\hat{\mu}}^f(1\!+\!\gamma_{\mu})U_{x,\mu}^{\dag}\psi_{x}^f
 \right\} \right],
\label{eqn:wilson}
\eea
where $\beta=6/g^2$ with $g$ the bare gauge coupling constant.%
\footnote{
Many flavor QCD using the staggered fermion formalism
is studied in Ref.~\citen{Columbia92}.
}
The index $f$ ($=1, \cdots, N_F$) is for the flavors.
We assume that all $N_F$ quarks are degenerate with the bare mass $m_0$, 
which is related to the hopping parameter $K$ by 
$ m_0 = (1/2a) (1/K-1/K_c) $, 
where $a$ is the lattice spacing and $K_c=1/8$ is the 
point where the bare mass vanishes.
The point $K=0$ corresponds to infinite quark mass.

In the formalism of Wilson quarks on the lattice, 
the flavor symmetry as well as C, P and T symmetries are exactly 
satisfied on the lattice with a finite $a$. 
However, chiral symmetry is explicitly broken 
even at $m_0=0$.
[For details about the effects of broken chiral symmetry, 
see Ref.~\citen{Stand26} and references cited there.]
However, through a perturbative study of Ward identities using 
the Wilson fermion action \cite{Bo},
it is shown that the effects of the broken chiral symmetry 
can be absorbed by appropriate renormalizations, 
including an additive renormalization of the quark mass,
i.e.\ by a shift of $K_c$ as a function of $\beta$.

We define the current quark mass in terms of an axial vector
Ward identity:
\be
2 m_q \langle\,0\,|\,P\,|\,\pi(\vec{p}=0)\,\rangle
= - m_\pi \, \langle\,0\,|\,A_4\,|\,\pi(\vec{p}=0)\,\rangle
\label{eq:mq}
\ee
where $P$ is the pseudoscalar density, $A_4$ the fourth component of the
local axial vector current, and
$m_\pi$ the pion (screening) mass \cite{ItohNP,Bo}. 
A multiplicative renormalization factor for the axial current,
which is not important in this study, is absorbed into the definition 
of the quark mass.
Numerical simulations show that the value of $m_q$ does not depend on
whether the system is in the high or the low temperature phase
when $\beta$ is large; $\beta \simm{>} 5.5$ for $N_F=2$ \cite{Tsukuba91}.

We define the chiral limit $K_c$ by the condition $m_q=0$.
Numerical simulations show that, for small $N_F$, $K_c(\beta)$ 
forms a smooth curve connecting 1/8 at $\beta = \infty$ 
and 1/4 at $\beta =0$.
When we further define the pion decay constant $f_\pi$ by 
$ %%\be
\langle\,0\,|\,A_4\,|\,\pi(\vec{p}=0)\,\rangle = m_\pi f_\pi,
$ %%\ee
either $m_\pi=0$ or $f_\pi=0$ is satisfied in the chiral limit.
The chiral symmetry is restored when $f_\pi=0$ in the chiral limit,
while $m_\pi=0$ if the chiral symmetry is spontaneously broken.

For QCD with a small $N_F$, the deconfinement transition 
is expected at high temperatures \cite{KanayaYKIS97}.
Therefore, we have to study the temperature dependence of
the results to distinguish between the deconfinement transition 
due to high temperatures 
and a bulk transition due to the effects of many flavors.
On a lattice with $N_t$ sites in the Euclidian time direction,
the temperature is given by $T=1/N_t a$.
When the beta function is negative, as in the case of small $N_F$,
the lattice spacing $a$ is a decreasing function of $\beta=6/g^2$.
In this case, $T$ increases with increasing $\beta$ 
for a fixed finite $N_t$.
On the other hand, when the beta function is positive, 
$T$ decreases with increasing $\beta$.
Therefore, in order to 
distinguish between the deconfinement transition 
due to high temperatures and a bulk transition
for general $N_F$, we have to vary $N_t$.

We perform simulations on lattices $8^2 \times 10 \times N_t$ 
($N_t =4$, 6, and 8), $16^2 \times 24 \times N_t$ ($N_t=16$)
and $18^2 \times 24 \times N_t$ ($N_t=18$).
We vary $N_F$ from 2 to 300.
For each $N_F$, we study the phase structure in the coupling 
parameter space $(\beta,K)$.
We adopt an anti-periodic boundary condition for quarks 
in the $t$ direction and
periodic boundary conditions otherwise.
We use the hybrid R algorithm \cite{Ralgo} for the generation of 
gauge configurations.%
\footnote{
The R algorithm has discretization errors of $O(N_F \Delta\tau^2)$ 
for the step size $\Delta\tau$ of a molecular dynamic evolution. 
As $N_F$ increases we have to decrease $\Delta\tau$, such as
$\Delta\tau$ =0.0025 for $N_F=240$, to reduce the errors.
We check that the errors are sufficiently small 
with our choices of $\Delta\tau$ for typical cases.
}
Statistical errors are estimated by the jack-knife method.

It should be noted that, in QCD with dynamical quarks, 
there are no order parameters for quark confinement. 
We discuss about confinement by comparing the screening pion mass 
and the lowest Matsubara frequency, and, simultaneously, consulting
the values of plaquette and the Polyakov loop.
In the followings, we call the pion screening mass simply 
the pion mass, and similarly for the quark mass.

\section{Strong coupling limit}
\label{sec:strong}

\begin{figure}[tb]
\centerline{
a)\epsfxsize=6.1cm\epsfbox{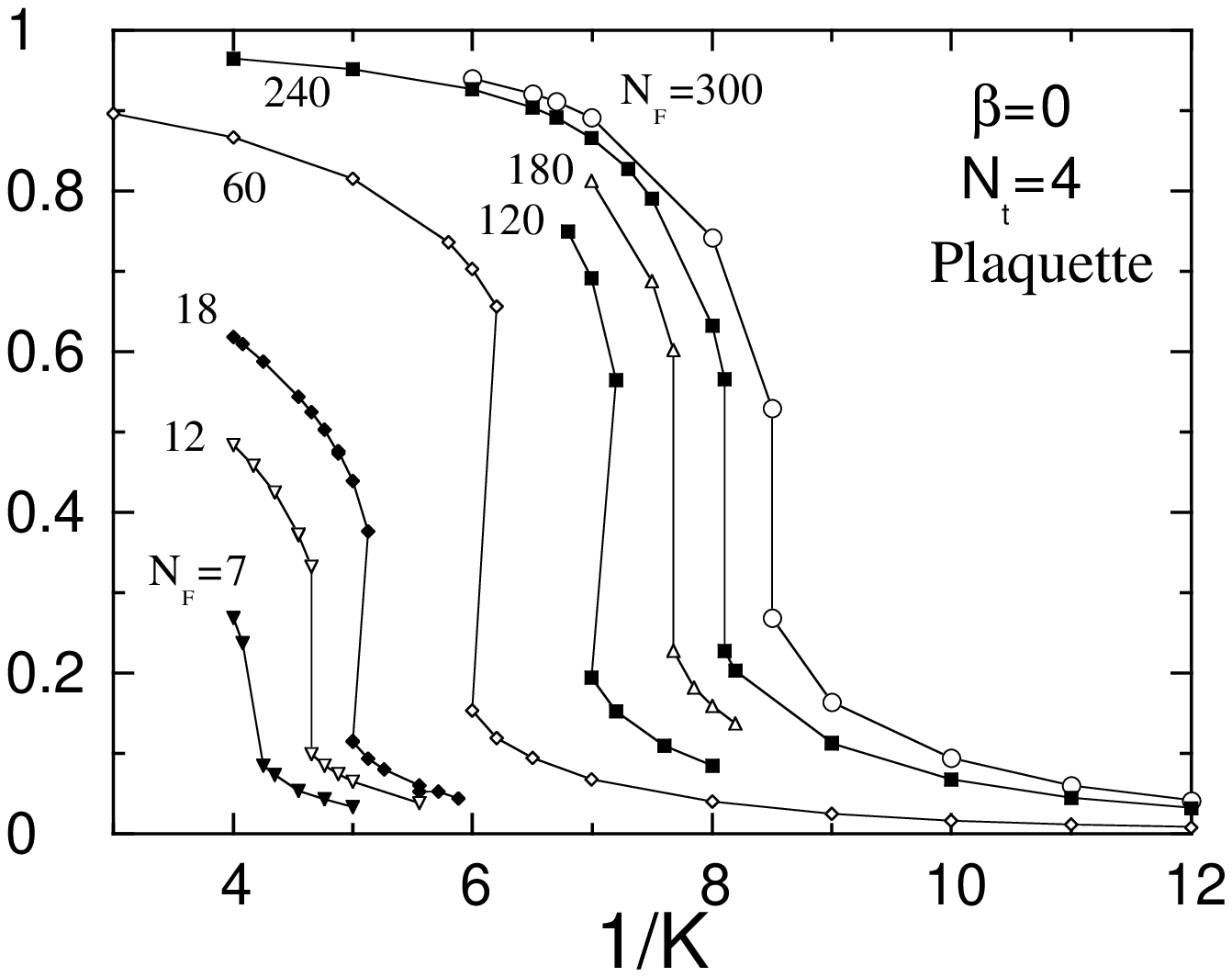}
\makebox[5mm]{}
b)\epsfxsize=6cm\epsfbox{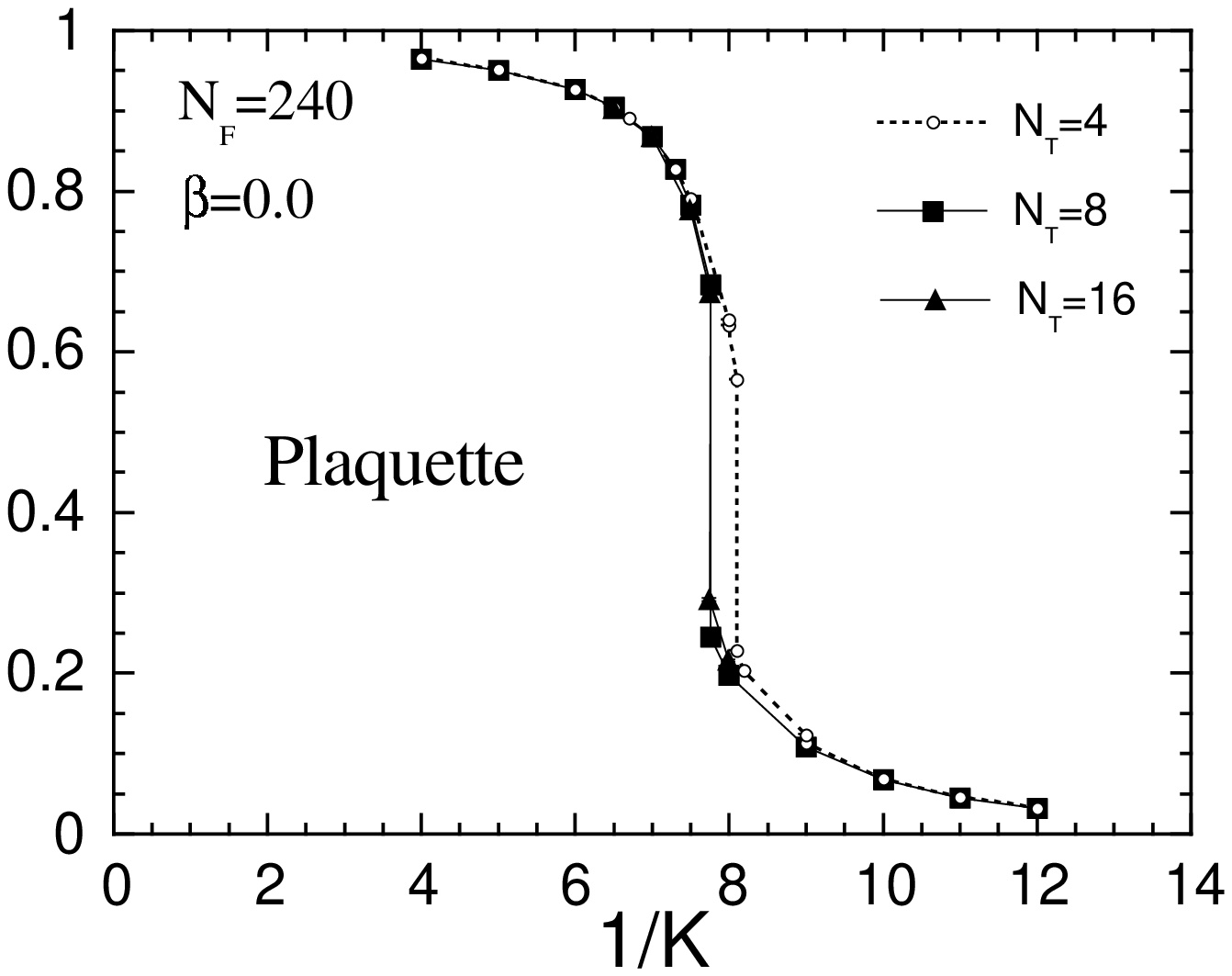}
}
\caption{
Plaquette at $\beta=0$ as a function of $1/K$.
(a) For various $N_F$ at $N_t=4$.
(b) For $N_F=240$ at $N_t=4$, 8 and 16.
$1/K_d \simeq 8.1(1)$ for $N_t=4$ and $1/K_d = 7.8(2)$ for $N_t=8$ 
and 16.
}
\label{B0.W}
\end{figure}

We first study the strong coupling limit, $\beta=0$,
where we expect the strongest confining force.
In a previous work \cite{previo}, we have shown that, when $N_F \ge 7$, 
light quarks are deconfined and chiral symmetry is restored 
at zero temperature even in the strong coupling limit:
For $N_F \leq 6$, we have only one confining phase from the heavy
quark limit $K=0$ up to the chiral limit $K_c=0.25$.
On the other hand, 
for $N_F \geq 7$, we find a strong first order transition at 
$K=K_d < K_c$.
When quarks are heavy ($K < K_d$), both plaquette and the Polyakov 
loop are small, and $m_\pi$ satisfies the PCAC relation 
$m_\pi^2 \propto m_q$. 
Therefore, quarks are confined and the chiral 
symmetry is spontaneously broken in this phase.
We find that $m_q$ in the confining phase is non-zero at the 
transition point $K_d$, 
i.e., the chiral limit does not belong to this phase.
On the other hand, when quarks are light ($K > K_d$), plaquette and 
the Polyakov loop are large.
In this phase, $m_\pi$ remains large in the chiral limit
and is almost equal to twice the lowest Matsubara frequency $\pi/N_t$. 
This implies that the pion state is a free two-quark state 
and, therefore, quarks are not confined in this phase.
The pion mass is nearly equal to the scalar meson mass, 
and the rho meson mass to the axial vector meson mass. 
The chiral symmetry is also manifest within
corrections due to finite lattice spacing.

\begin{figure}[tb]
\centerline{
\epsfxsize=7cm \epsfbox{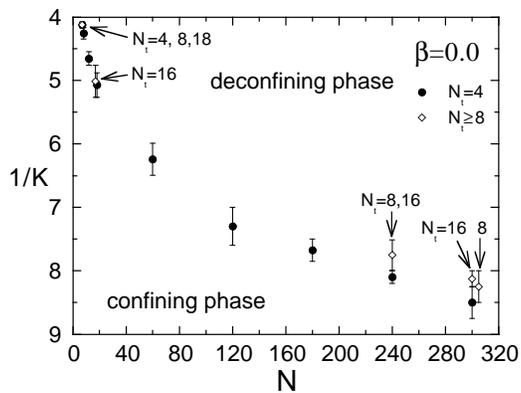}
}
\vspace{-0.3cm}
\caption{
The transition point $1/K_d$ at $\beta=0$ versus $N_F$ 
for $N_t=4$ and $N_t\ge 8$. 
For clarity, data at $N_t=8$ for $N_F=300$ is slightly shifted 
to a larger $N_F$ in the figure. 
}
\label{Kd.Nf}
\end{figure}

We now extend the study to larger $N_F$.
In Fig.~\ref{B0.W}(a), we show the results of plaquette at $N_t=4$
for $N_F =7$ -- 300.
Clear first order transition can be seen at $K$ below 0.25.
We then study the $N_t$ dependence of the results, 
as shown in Fig.~\ref{B0.W}(b) for the case of $N_F=240$.
Results for $N_F=7$ is given in Ref.~\citen{previo}.
We find that,
although the transition point shows a slight shift to smaller $1/K$
when we decrease $T=1/N_t a$ from $N_t=4$ to 8, it stays at 
the same point for $N_t \geq 8$.
The chiral limit, $1/K=4$, cannot be achieved in the confining phase 
even at $T=0$. 
Therefore, we conclude that the transition is a bulk transition 
(transition that exists at zero-temperature).
From these results, we obtain the phase diagram at $\beta=0$
shown in Fig.~\ref{Kd.Nf}.

\begin{figure}[tb]
\centerline{
a)\epsfxsize=6.3cm\epsfbox{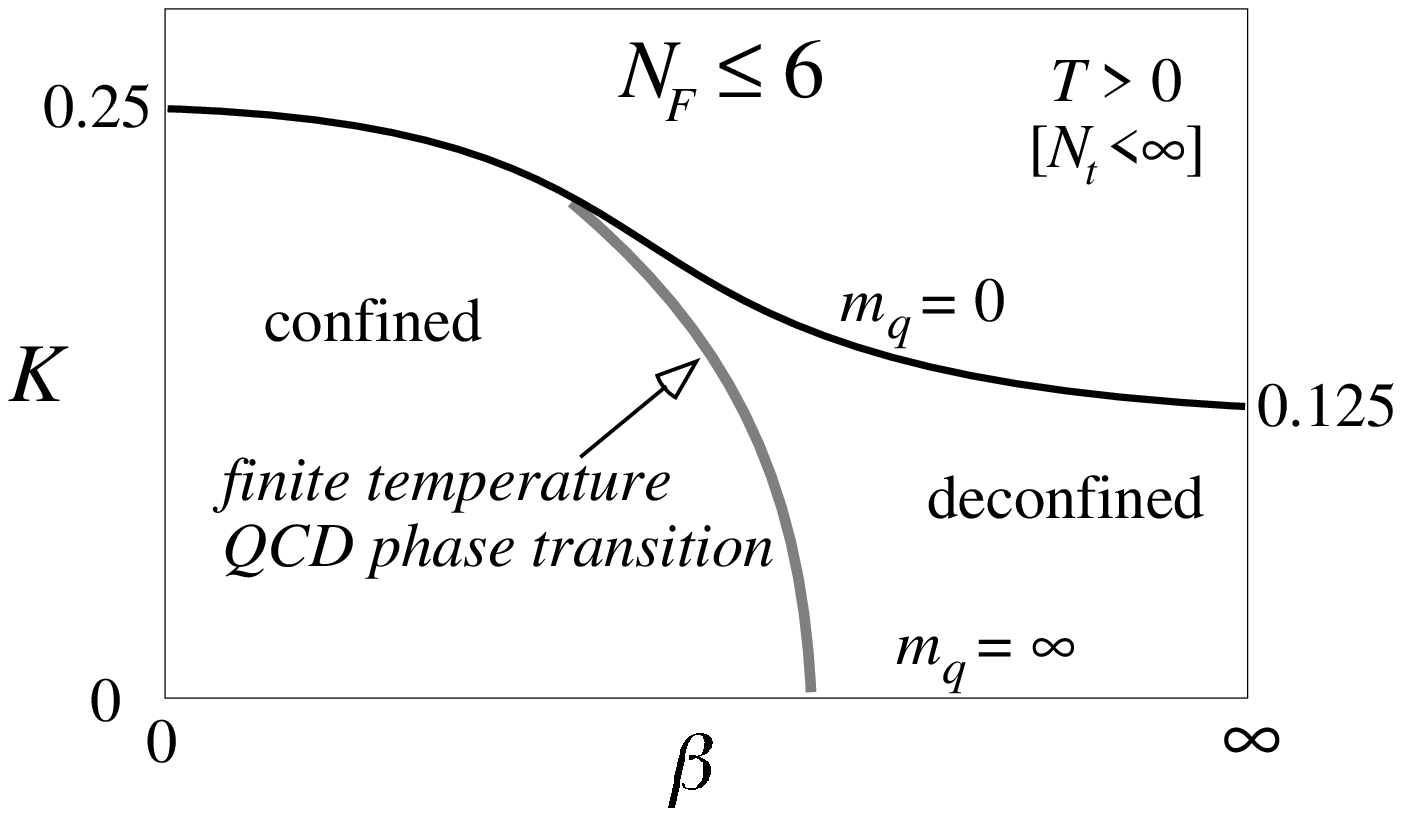}
\makebox[3mm]{}
b)\epsfxsize=6.3cm\epsfbox{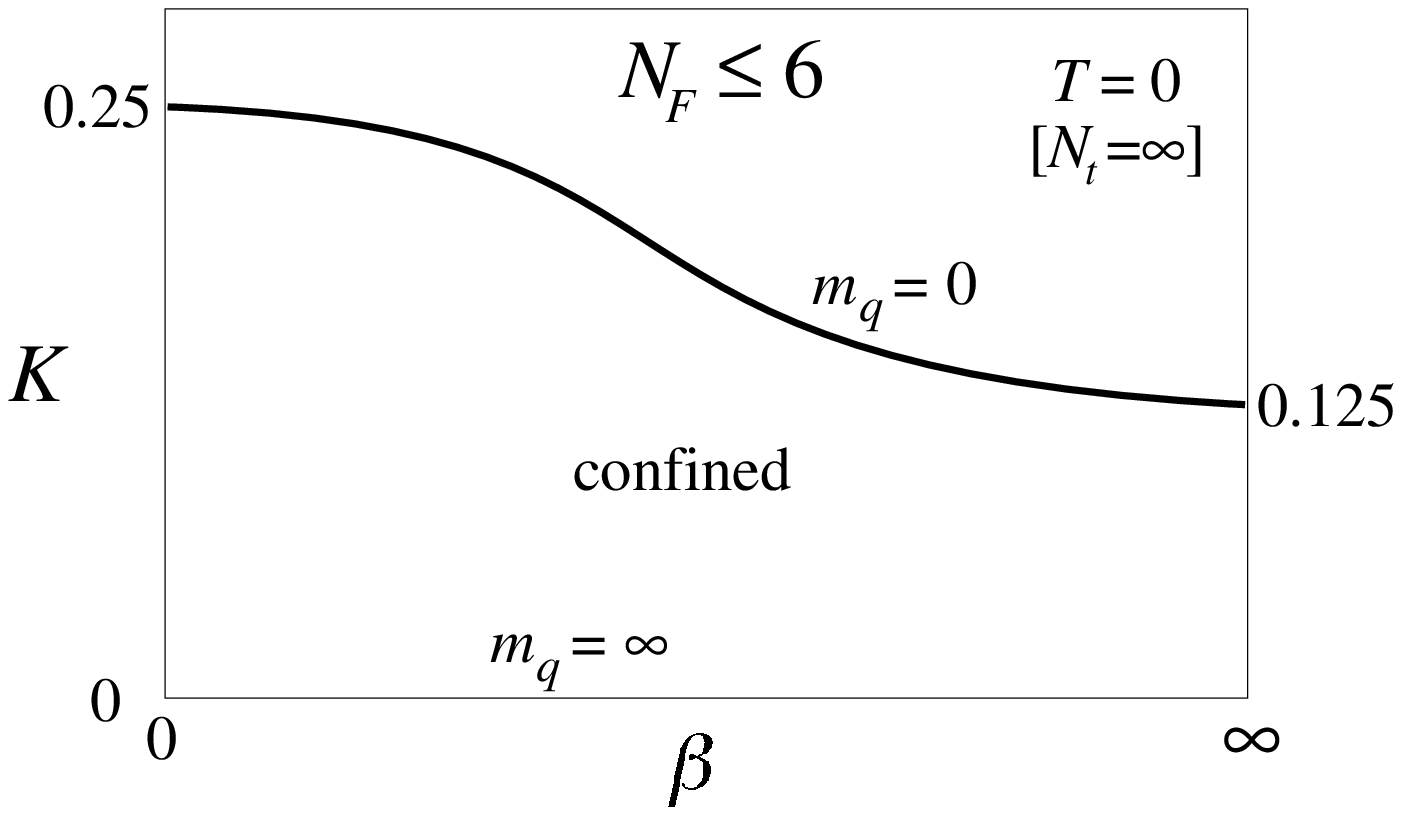}
}
\caption{The phase structure for $N_F \leq 6$;
(a) at finite temperatures, and
(b) at zero temperature.
}
\label{NfSmall}
\end{figure}

\section{Phase structure at $\beta > 0$}
\label{sec:finitebeta}

Let us now study the case $\beta > 0$.
On a lattice with a fixed finite $N_t$, we have the 
finite temperature deconfining transition at finite $\beta$ 
when the quark mass is sufficiently heavy ($K \approx 0$).
For $N_F \leq 6$, our previous study shows that the finite 
temperature transition line crosses the $K_c$ line at finite $\beta$
\cite{Stand26}.
A schematic diagram of the phase structure for this case is shown 
in Fig.~\ref{NfSmall}.%
\footnote{
See Ref.~\citen{Stand26} and references cited there 
for discussions about complexities due to the lack of chiral 
symmetry with Wilson fermions.
}
The location of the finite temperature transition line moves towards 
a larger $\beta$ as $N_t$ is increased.
In the limit $N_t=\infty$, 
the finite temperature transition line will shift to $\beta=\infty$
so that only the confining phase is realized for $N_F \leq 6$
at $T=0$.  See Fig.~\ref{NfSmall}(b).
There, we have a monotonous RG flow towards the strong coupling
limit on the $K_c$ line.

\begin{figure}[tb]
\centerline{
a)\epsfxsize=6.2cm\epsfbox{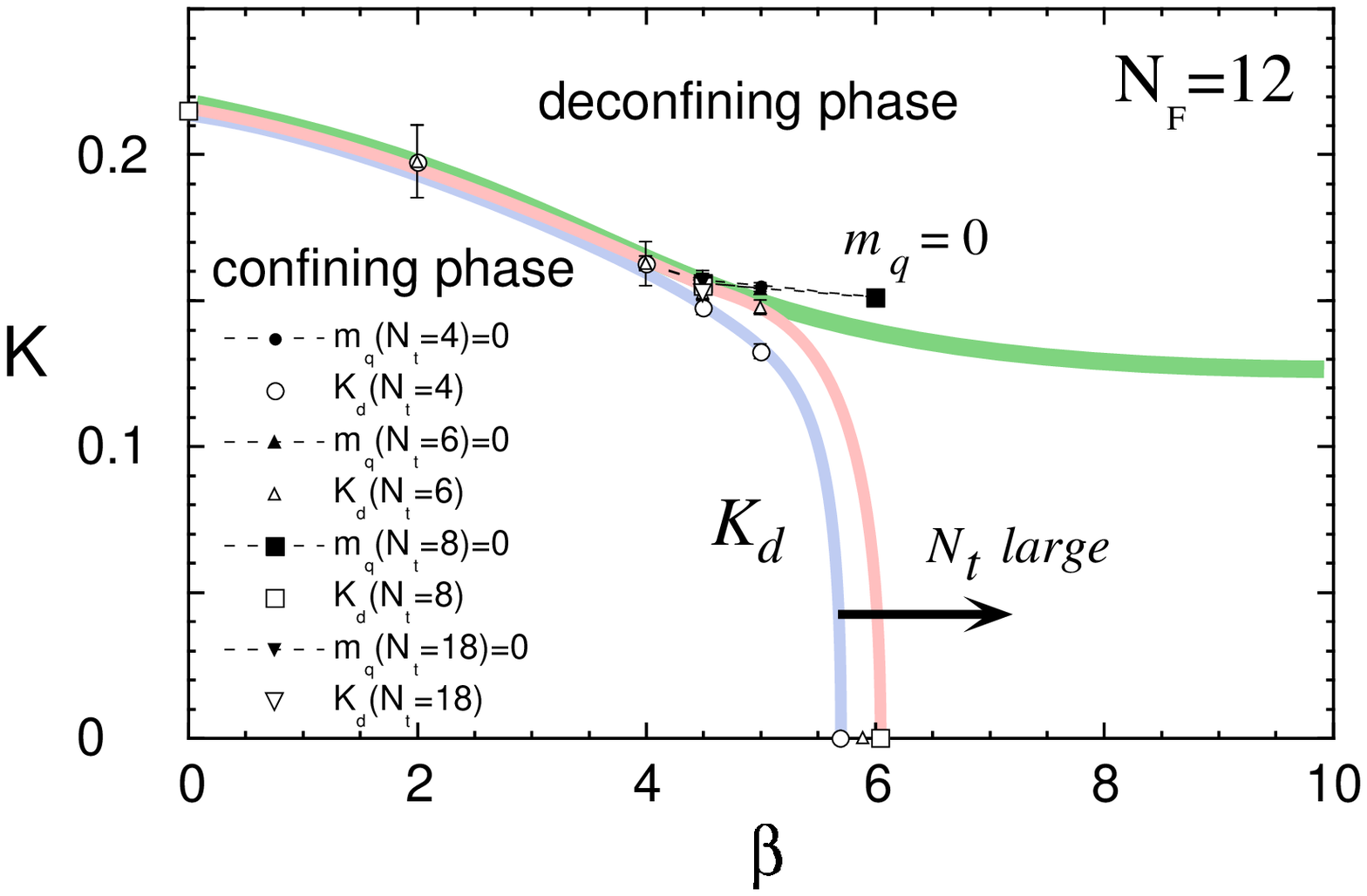}
\makebox[3mm]{}
b)\epsfxsize=6.2cm\epsfbox{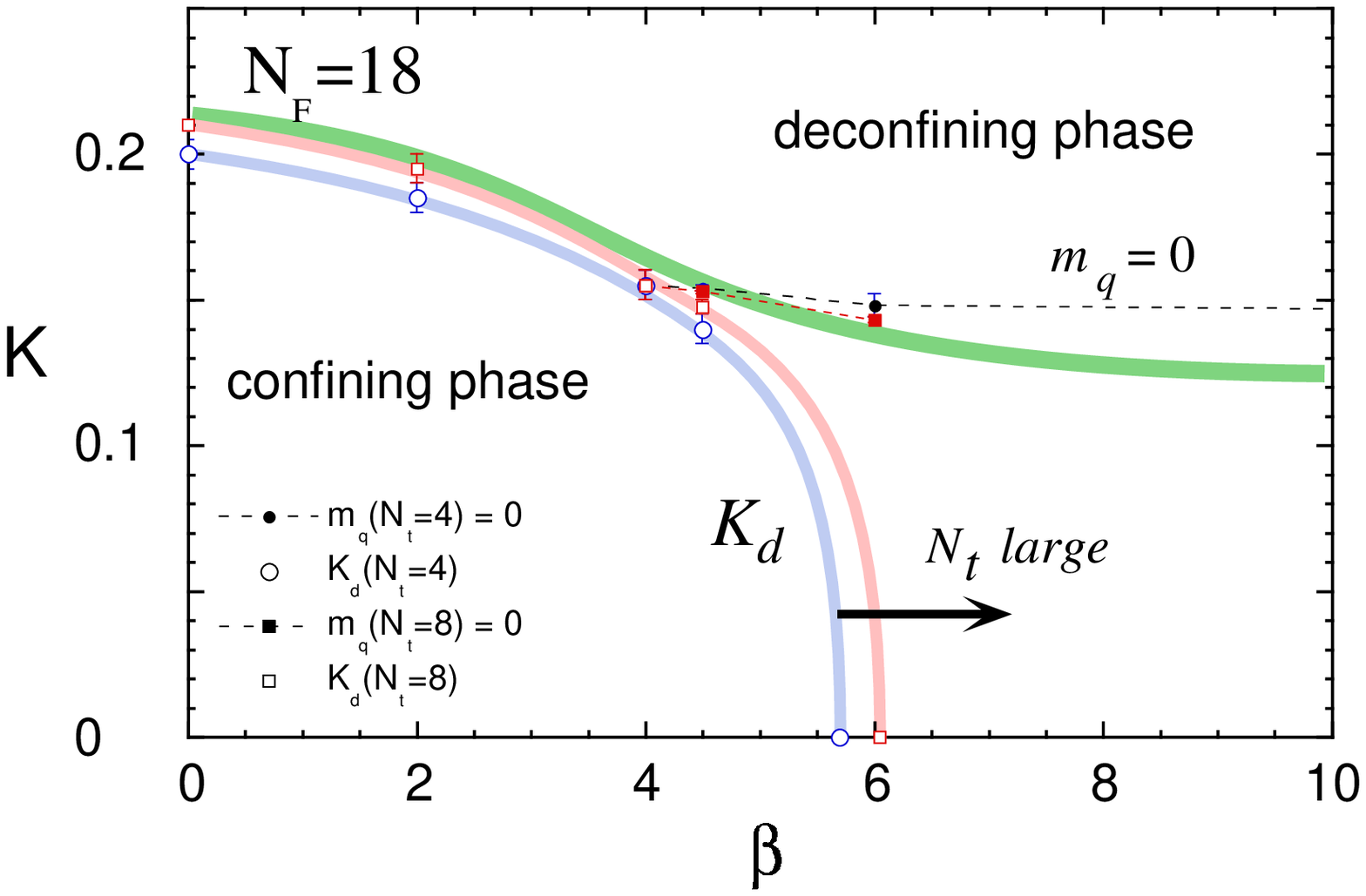}
}
\caption{
Phase diagram for (a) $N_F=12$ and (b) $N_F=18$.
Dark shaded lines represent our conjecture for the bulk transition 
line in the limit $N_t=\infty$.
}
\label{Nf12_18}
\end{figure}

\begin{figure}[tb]
\centerline{
\epsfxsize=6.2cm\epsfbox{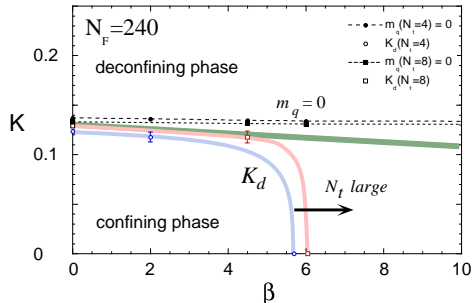}
}
\vspace{-0.3cm}
\caption{Phase diagram for $N_F=240$.}
\label{Nf240}
\end{figure}

When $N_F \geq 7$, we have seen that there exist a bulk 
deconfining transition in the strong coupling limit.
We have to clarify the relation between
the finite temperature transition at small $K$ 
and the bulk transition at $\beta=0$. % for large $N_F$.

At $\beta=0$, the location of the bulk deconfining transition $K_d$
moves toward larger $K$ with decreasing $N_F$.
From Fig.~\ref{Kd.Nf}, we find 
$1/K_d \simeq 8.5$, 8, 7.6, 7.2, 6.1, and 5.0 for
$N_F= 300$, 240, 180, 120, 60, and 18, respectively, at $\beta=0$.
Recall that $1/K=8$ is the point for massless free quarks.

We perform simulations at $\beta \approx 0.0$ -- 6.0
varying $N_F$ from 2 to 300.
From the numerical data, we obtain the phase diagrams
sown in Fig.~\ref{Nf12_18} for $N_F=12$ and 18, and 
in Fig.~\ref{Nf240} for $N_F=240$.
When $N_t =4$ or 8, the phase boundary line between confining
and deconfining phase bends down at finite $\beta$
due to the finite temperature phase transition
of the confining phase.
The dark shaded lines are our conjecture for the phase boundary between 
the confining phase and the deconfining phase at zero temperature
($N_t=\infty$).

In Figs.~\ref{Nf12_18} and \ref{Nf240}, 
the dashed lines in the deconfining phase represents the 
chiral limit, $m_q=0$. 
This line also corresponds to the minimum point of $m_\pi^2$.
We have checked that, for the case $N_F=240$, the quark propagator 
in the Landau gauge actually shows the chiral symmetry, 
$\gamma_5 G(z) \gamma_5 = - G(z)$, on the $m_q=0$ line. 
For $N_F \simm{>} 240$, we find that the $m_q=0$ line in the 
deconfining phase, which starts
at $1/K=8$ in the weak coupling limit $\beta=\infty$, 
reaches the strong coupling limit,
as shown in Fig.~\ref{Nf240} for $N_F=240$.
For a smaller $N_F \simm{<} 100$, 
because the bulk transition line $K_d$ in the strong coupling 
region shifts toward larger $K$ with decreasing $N_F$, 
the $m_q=0$ line in the deconfining phase
hits the $K_d$ line at finite $\beta$.
For example, in the case $N_F=18$, it hits at $\beta = 4.0$ -- 4.5,
as shown in Fig.~\ref{Nf12_18}(b).

\section{Nature of the deconfining phase for large $N_F$}
\label{sec:deconf}

In order to study the nature of the deconfining phase, 
we first study the case of large $N_F$, 
because the range of the deconfining phase is
wider at larger $N_F$ (see Fig.~\ref{Kd.Nf}) and therefore 
the parameter dependence of physical quantities becomes easier to study.
In particular, for $N_F \simm{>} 240$, we are able to study 
the chiral limit in the deconfining phase at any values of $\beta$
(see Fig.~\ref{Nf240}).
Because of this, we first intensively investigate the cases $N_F=240$
and 300, and then decrease $N_F$.

\subsection{$N_F=240$ and $300$}

\begin{figure}[t]
\centerline{
a)\epsfxsize=7.1cm\epsfbox{figure/nf240.mass.ps.2}
\makebox[1mm]{}
b)\epsfxsize=5.4cm\epsfbox{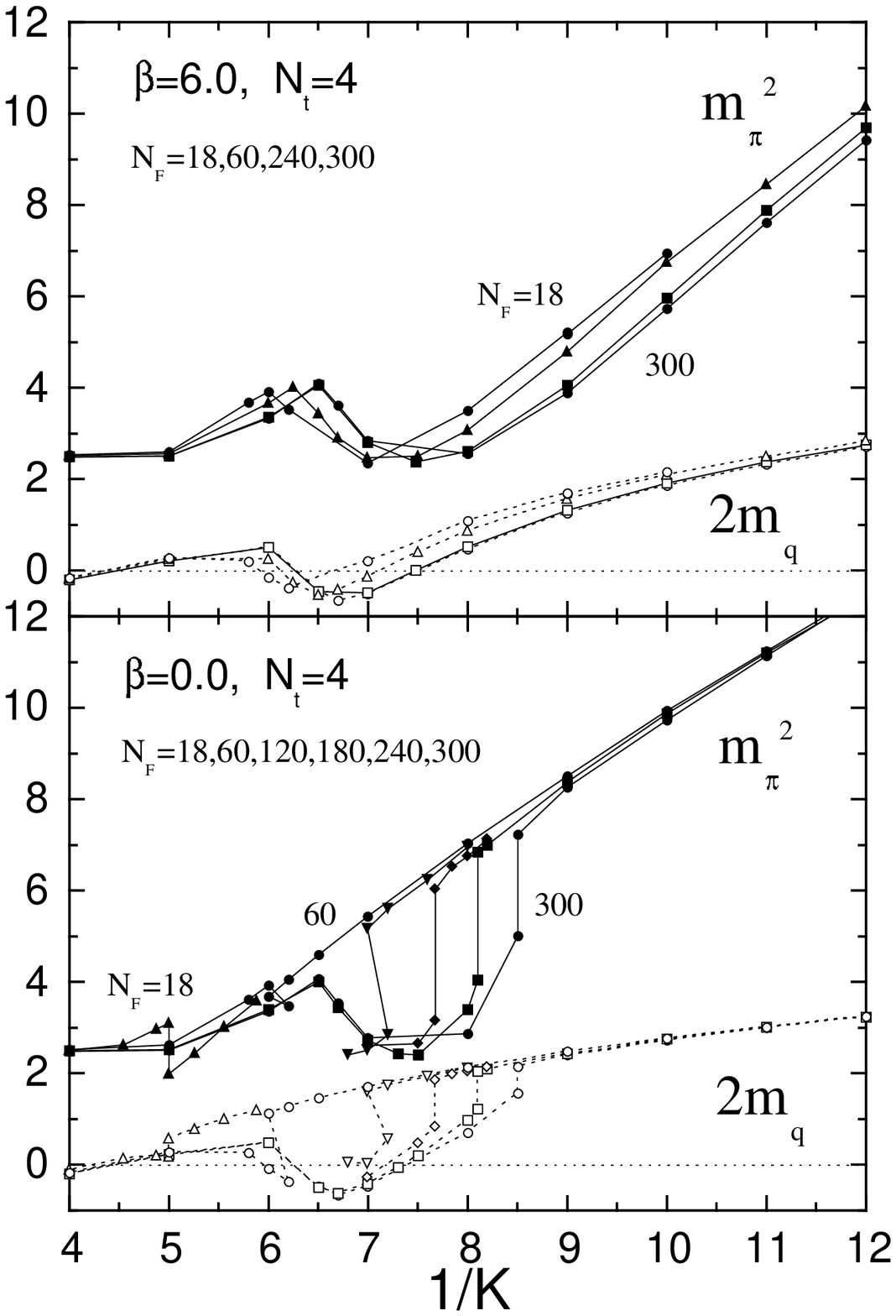}
}
\caption{
Results of $m_\pi^2$ and 2$m_q$ versus $1/K$ at $N_t=4$:
(a) $N_F=240$ at various $\beta$.
(b) $N_F=18$ -- 300 at $\beta=6.0$ and 0.0.
}
\label{massNfLarge}
\end{figure}

Fig.~\ref{massNfLarge}(a)
shows the results of $m_\pi^2$ and $2m_q$ in the deconfining 
phase for $N_F=240$ 
at $\beta=0.0$, 2.0, 4.5, 6.0, and 100.0 on the $N_t=4$ lattice.
A very striking fact is that the shape of $m_\pi^2$ and $2m_q$ 
as a function of $1/K$
only slightly changes for $1/K < 8$
when the value of $\beta$ decreases from $\infty$
down 0. 
[Note that $\beta=\infty$ corresponds to the case of free quarks.
Due to the doublers, $m_q$ defined by (\ref{eq:mq}) shows a 
non-trivial $1/K$ dependence at $1/K < 8$ also for a free quark.]
Only the position of the local minimum of $m_\pi^2$ at $1/K\simeq8$, 
which corresponds to the vanishing point of $m_q$, 
slightly shifts toward smaller $1/K$.
We obtain similar results also for $N_t=8$.
This suggests that quarks are almost free down to $\beta=0$ 
in the deconfining phase.

The results for $N_F=300$ are essentially the same as those
for $N_F=240$, except for very small shifts of the transition point
and the minimum point of $m_\pi^2$.

\subsection{Renormalization group flow for $N_F=240$}

From the perturbation theory, 
the $m_q=0$ point at $\beta=\infty$ is a trivial IR fixed point
for $N_F \ge 17$. 
The phase diagram shown in Fig.~\ref{Nf240} suggests that 
there are no other fixed points on the $m_q=0$ line at finite $\beta$.
In order to confirm this, we investigate 
the direction of the renormalization group (RG) flow along the $m_q=0$ 
line for $N_F=240$, 
using a Monte Carlo renormalization group (MCRG) method.

We make a block transformation for a change of scale factor 2,
and estimate the quantity $\Delta \beta = \beta(2a) - \beta(a)$:
We generate configurations
on an $8^4$ lattice on the $m_q=0$ points at $\beta=0$ and 6.0
and make twice blockings. 
We also generate configurations on a $4^4$ lattice 
and make once a blocking.
Then we calculate $\Delta\beta$ 
by matching the value of the plaquette at each step.%
\footnote{
It is known for the pure SU(3) gauge theory, 
that one has to make a more careful analysis 
using several types of Wilson loop with many blocking steps 
to extract a precise value of $\Delta \beta$.
We reserve elaboration of this point and a fine tuning of $1/K$ at each
$\beta$ for future works.
For $N_F=240$, because the velocity of the RG flow is large, 
we will be able to obtain the sign and an approximate value of
$\Delta\beta$ by a simple matching.
}
From a matching using our data, we obtain $\Delta \beta \simeq 6.5$ at
$\beta=0$ and $10.5$ at $\beta=6.0$.
The value obtained from the two-loop perturbation theory is 
$\Delta \beta \simeq 8.8$ at $\beta = 6.0$.
The signs are the same and the magnitudes are comparable.
This suggests that the direction of the RG flow on the $m_q=0$ line
at $\beta=0$ and 6.0 is the same as that at $\beta=\infty$.
This further suggests that there are no fixed points at finite $\beta$.
All of the above imply that the theory is trivial for $N_F=240$.

\subsection{$240 \ge N_F \ge 17$}

Now we decrease $N_F$ from 240.
As discussed in Sec.~\ref{sec:finitebeta},
the area of the deconfining phase decreases with 
decreasing $N_F$ in the strong coupling region. 
However, except for this shift of the bulk transition point, 
the shape and the values of $m_\pi^2$ and $m_q$ are quite similar
when we vary $N_F$ from 300 down to 17.
Fig.~\ref{massNfLarge}(b) shows $m_\pi^2$ and $2m_q$ 
for $N_F=18$ -- 300 at $\beta=6.0$ and 0.
At $\beta=6.0$, the shapes of $m_\pi^2$ are almost identical to each 
other, except for a small shift toward smaller $1/K$ as $N_F$ is 
decreased.

These facts suggest that, for $N_F \geq 17$, the nature of physical 
quantities in the deconfining phase is almost identical to that 
observed for $N_F=240$ and 300, i.e., quarks are almost free.
Therefore, we conjecture that the direction of the RG flow at 
$\beta \simm{<} 6$ is identical to the case $N_F=240$.
Combining this with the perturbative result
that, for $N_F \ge 17$, $\beta=\infty$ is the IR fixed
point, we conjecture that the RG flow along the massless quark line 
in the deconfining phase is uniformly directing to smaller $\beta$,
as in the case of $N_F=240$.
When this is the case, the theory has only one IR fixed point at 
$\beta=\infty$ --- the theory is trivial for $N_F \ge 17$.

\subsection{$16 \geq N_F \geq 7$}
\label{sec:nf167}

As shown in Sec.~\ref{sec:strong},
quark confinement is lost for $N_F \ge 7$ at $\beta=0$.
We have intensively simulated the cases $N_F=12$ and 7.
The phase diagram for $N_F=12$ is shown in Fig.~\ref{Nf12_18}(a).
At $\beta \simm{<} 6.0$, the gross feature of the phase diagrams 
for $N_F=7$ and 12 is quite similar to the case $N_F=18$ shown 
in Fig.~\ref{Nf12_18}(b). 
Physical quantities at $\beta \simm{<} 6.0$ also show behavior 
similar to that shown in Fig.~\ref{massNfLarge} for $N_F \geq 17$.
Therefore, we consider it probable that the direction of the RG flow 
in the deconfining phase at small $\beta$ is towards a larger $\beta$ 
as in the case of $N_F \geq 17$.

On the other hand, when $N_F \le 16$, the theory is asymptotic free. 
Therefore, the RG flow at $\beta=\infty$ is opposite to that 
for $N_F \geq 17$. 
This means that we have an IR fixed point somewhere at a finite value 
of $\beta$.
The continuum limit is governed by this IR fixed point. 

On the $m_q=0$ line around $\beta=4.5$,
we find that $m_\pi$ is roughly twice the lowest Matsubara frequency.
This implies that the quarks are not confined and almost free. 
Therefore, the anomalous dimensions are small at the IR fixed point, 
suggesting that the IR fixed point locates at finite $\beta$.
In order to pin down the position of the IR fixed point,
we have to make a detailed study of RG flows in a wider coupling parameter 
space.
We reserve these studies for future works.

In summary, we conjecture that, for $16 \geq N_F \geq 7$,
we have an IR fixed point at finite $\beta$.
The theory in the continuum limit is a non-trivial theory with 
small anomalous dimensions, however, without confinement.

\begin{figure}[tb]
\centerline{
\epsfxsize=6cm\epsfbox{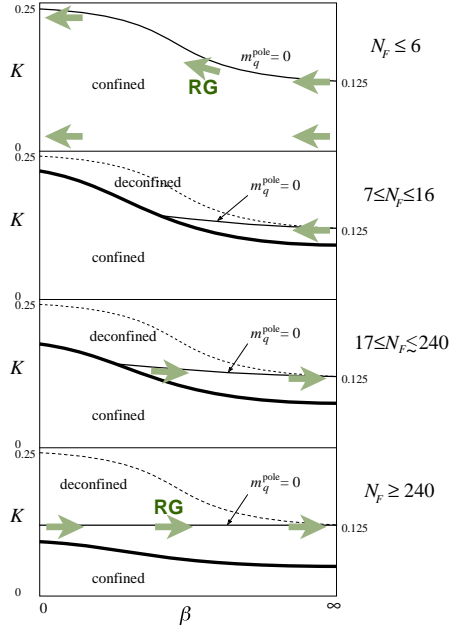}
}
\vspace{-0.3cm}
\caption{Phase structure and the RG flow at $T=0$
for general number of flavors.
Thin dashed lines in the phase diagrams for $N_F \geq 7$ represent 
the location of the chiral limit for $N_F \leq 6$ as a guide for eyes.}
\label{RGflow}
\end{figure}

\section{Conclusions}
\label{sec:conclusions}

Based on our numerical data combined with the results of 
the perturbation theory,
we propose Fig.~\ref{RGflow} for the phase structure and the RG flow 
at $T=0$ for general number of flavors. 

When $N_F \le 6$, 
quarks are confined for any values of the current quark mass and $\beta$.
The chiral limit $K_c(\beta)$, where the current quark mass $m_q$ vanishes, 
belongs to the confining phase.

When $N_F \ge 7$, there is no chiral limit in the confining phase.
There is a line of a first order phase transition from the confining phase
to a deconfining phase at a finite current quark mass for all values
of $\beta$. 
The chiral limit exists only in the deconfining phase.

In order to study the nature of the deconfining phase for large $N_F$, 
we first increase $N_F$, because the region of the deconfining phase 
becomes larger with larger $N_F$.
For $N_F=240$, a MCRG study along the massless quark line shows that 
the direction and the magnitude of the RG flow at small $\beta$ is 
consistent with the result of the perturbation theory, 
suggesting that the direction of the RG flow is identical for 
all values of $\beta$, as shown in Fig.~\ref{RGflow}.
Therefore, there is only one IR fixed point at $\beta=\infty$, 
so that the theory in the continuum limit is trivial.
We also find that the $K$-dependences of pion screening mass and current 
quark mass in the deconfining phase are almost identical to those of
free Wilson quarks.
Also for smaller values of $N_F \geq 7$, 
the general features of the phase diagram and the 
$K$-dependence of physical quantities 
are quite similar to the case of large $N_F \sim 240$.
Therefore, we conjecture that the direction of the RG flow is 
the same as in the case of $N_F=240$, 
at least for $\beta \simm{<} 6.0$ we simulate.

For $N_F \geq 17$, because the RG flow in the weak coupling limit 
has the same direction due to the lack of asymptotic freedom, 
we conjecture a uniform RG flow towards the trivial IR fixed point,
as in the case of $N_F\geq 240$.
Thus the theory is trivial for $N_F \ge 17$.

When $7 \le N_F \le 16$, because the theory is asymptotically free,
we expect a non-trivial IR fixed point at finite $\beta$.
In this case, the theory in the continuum limit is a non-trivial 
theory with anomalous dimensions, however, without confinement.

\section*{ACKNOWLEDGEMENTS}
\vspace*{-2mm}
Numerical simulations are performed with 
HITAC S820/80 at KEK, and Fujitsu VPP500/30 and
QCDPAX at the University of Tsukuba.
KK would like to thank the participants of YKIS'97 for valuable comments.
KK is also grateful to Robert D.\ Pisarski, Motoi Tachibana and 
Bengt Petersson for useful discussions.
This work is in part supported by the Grants-in-Aid of Ministry of 
Education, Science and Culture (Nos.~08NP0101 and 09304029).


\begin{thebibliography}{99}

\bibitem{lat94}
Y.\ Iwasaki, K.\ Kanaya, S.\ Sakai, and T.\ Yoshi\'e,
Nucl.\ Phys.\ {\bf B}(Proc.\ Suppl.){\bf 42} (1995), 502.

\bibitem{lat96} 
Y.\ Iwasaki, K.\ Kanaya, S.\ Kaya, S.\ Sakai, and T.\ Yoshi\'e,
Nucl.\ Phys.\ {\bf B}(Proc.\ Suppl.){\bf 53} (1997), 449.

\bibitem{Banks1} 
T.\ Banks and A.\ Zaks, \NP{B196,1982,173}.

\bibitem{Oehme}
R.\ Oehme and W.\ Zimmermann, \PR{D21,1980,471}, 1661.
R.\ Oehme, \PR{D42,1990,4209}.

\bibitem{Nishijima}
K.\ Nishijima, \NP{B238,1984,601}; \PTP{75,1986,22};
{\it ibid.} \andvol{77,1987,1053}.

\bibitem{Appelquist}
T.\ Appelquist et al., \PRL{77,1996,1214}.

\bibitem{sigma}
R.D.\ Pisarski and D.L.\ Stein, \PR{B23,1981,3549};
\JP{A14,1981,3341}.
A.J.\ Paterson, \NP{B190[FS3],1981,188}.

\bibitem{Kogut79}
J.\ Kogut, R.\ Pearson and J.\ Shigemitsu, \PRL{43,1979,484}.

\bibitem{SCENc}
H.\ Kluberg-Stern et al., %, A.\ Morel, O.\ Napoly and B.\ Petersson,
\NP{B190[FS3],1981,504}.
N.\ Kawamoto, \NP{B190[FS3],1981,617}.
N.\ Kawamoto and J.\ Sumit, \NP{B192,1981,100}.

\bibitem{SCEmf}
J.\ Smit, \NP{B175,1980,307}.
J.\ Greensite and J.\ Primack, \NP{B180[FS2],1981,170}.
H.\ Kluberg-Stern, A.\ Morel and B.\ Petersson,
\NP{B215[FS7],1983,527}.

\bibitem{SCEmfMq}
J.-M.\ Blairon et al., %, R.\ Brout, F.\ Englert and J.\ Greensite,
\NP{B180[FS2],1981,439}.

\bibitem{SCEMq}
K.\ Wilson, in ``New phenomena in subnuclear physics'' (Erice 1975),
ed.\ A.\ Zichichi (Plenum, NY, 1977).


\bibitem{previo} 
Y.\ Iwasaki, K.\ Kanaya, S.\ Sakai, and T.\ Yoshi\'e,
\PRL{69,1992,21}.


\bibitem{KanayaYKIS97}
K.\ Kanaya, in these proceedings.

\bibitem{Columbia92}
M.\ Fukugita, S.\ Ohta and A.\ Ukawa, \PRL{60,1988,178}.
J.B.\ Kogut and D.K.\ Sinclair, \NP{B295[FS21],1988,465}.
S.\ Ohta and S.\ Kim, \PR{D44,1991,504}.
F.\ Brown et al., \PR{D46,1992,5655}.
P.H.\ Damgaard et al., %, U.M.\ Heller, A.\ Krasnitz and P.\ Olesen,
\PL{B400,1997,169}.

\bibitem{Stand26} 
Y.\ Iwasaki, K.\ Kanaya, S.\ Kaya, S.\ Sakai, and T.\ Yoshi\'e,
\PR{D54,1996,7010}.

\bibitem{Bo}
M.\ Bochicchio et al., \NP{B262,1985,331}.

\bibitem{ItohNP}
S.\ Itoh, Y.\ Iwasaki, Y.\ Oyanagi and T.\ Yoshi\'e,
\NP{B274,1986,33}.

\bibitem{Ralgo}
S.\ Gottlieb et al., %W.\ Liu, D.\ Toussaint, R.L.\ Renken and R.L.\ Sugar,
\PR{D35,1987,2531}.

\bibitem{Tsukuba91}
Y.\ Iwasaki, K.\ Kanaya, S.\ Sakai and T.\ Yoshi\'e,
\PRL{67,1991,1494}.

\end{thebibliography}
\end{document}